\documentclass[11pt]{article}

\usepackage[letterpaper,margin=1in]{geometry}
\usepackage[T1]{fontenc}
\usepackage[utf8]{inputenc}
\IfFileExists{lmodern.sty}{\usepackage{lmodern}}{}
\usepackage{textcomp}
\usepackage{amsmath,amssymb}
\usepackage{graphicx}
\usepackage{booktabs}
\usepackage{array}
\usepackage{longtable}
\usepackage{caption}
\IfFileExists{lmodern.sty}{\usepackage{microtype}}{\usepackage[expansion=false]{microtype}}
\usepackage{xcolor}
\usepackage{gensymb}
\usepackage[hidelinks]{hyperref}
\IfFileExists{orcidlink.sty}{\usepackage{orcidlink}}{}

\graphicspath{{figures/}}
\providecommand{\tightlist}{\setlength{\itemsep}{0pt}\setlength{\parskip}{0pt}}

\title{\vspace{-2em}\bfseries Routing LLM Inference to the Cleanest Grid in Real Time\\[0.4em]
\large\mdseries A marginal-emissions-driven carbon-aware fabric: live multi-region feasibility on real GPUs, a telemetry-calibrated energy-and-carbon methodology, and a year-long marginal-emissions replay}

\author{
Aleks Bernhard\thanks{\texttt{aleks@solyx.ai}} \and Arif Baran Yardimci\thanks{\texttt{aby23@cornell.edu}}\\[0.3em]
\normalsize Solyx AI, Inc., Miami, FL, USA}

\date{}

\begin{document}
\maketitle
\thispagestyle{empty}

\begin{abstract}

Large-language-model (LLM) inference is now a material and fast-growing electricity load, and the marginal carbon intensity of that load varies by more than an order of magnitude across grid regions and across the day. Routing inference requests toward cleaner regions is therefore an attractive, low-friction lever --- no model retraining, no hardware change, only a placement decision. We report a live validation of carbon-aware inference routing on real multi-region GPU testbeds, driven by WattTime marginal operating emissions rate (MOER) signals, with three properties uncommon in prior work: (1) a realistic \emph{blind} baseline that is the actual production pressure-based router rather than uniform placement; (2) per-request energy attributed from GPU telemetry (NVIDIA DCGM) via measured concurrency curves rather than nameplate TDP; and (3) carbon settlement of every request against WattTime\textquotesingle s \emph{historical} MOER, not only the forecast that drove the routing decision.

The central result is one of feasibility: a MOER signal can steer live inference workloads across a multi-region GPU testbed, with no observed dispatch failures, as a strict and reversible overlay on the existing production router. Over multi-day runs on live two-region testbeds, the signal measurably moved real inference load between regions and settled emissions in the intended direction --- establishing that the mechanism works end-to-end on production hardware. What a short, two-site live run \emph{cannot} establish is the carbon magnitude --- how much this saves across grids and seasons --- which we quantify with a year-long historical-MOER replay across a grid-diverse CONUS fleet (\S6). The replay indicates a substantial potential reduction, from roughly 23\% under a soft multiplier policy matching the live production overlay\textquotesingle s form up to 34--51\% under an aggressive cleanest-first dispatch and alternative treatments of the zero-MOER curtailment intervals. In the primary replay configuration, carbon-aware placement reduces modeled GPU-attributable operational emissions by a central 50.9\% versus round-robin (51.0\% versus a modeled carbon-blind load balancer) (95\% block-bootstrap CI 48.5--53.3\%, \emph{conditional on the modeled fleet, routing policy, workload and zero-MOER-credit assumptions}). Because the replay dispatches directly against historical MOER rather than a forecast, these reductions represent an upper-bound opportunity under the modeled configuration; forecast error would reduce operationally realized savings. In the freely-placeable decomposition before session pinning, hourly routing that seeks the lowest MOER contributes approximately 22.4 percentage points --- roughly 40\% of the 54.0\% placement reduction --- beyond the static annual-mean-MOER policy. These are modeled results for one specified fleet and historical year, not a universal estimate; \S6 states the configuration and its assumptions. We also record a practical observation: when \emph{comparing} regions, rank by absolute MOER rather than the percentile \emph{signal-index}, which is normalized within each region and answers a \emph{temporal}, not a spatial, question. With potential magnitude characterized under the replay assumptions (\S6), the remaining work adds a physical capacity/headroom constraint from operational curtailment data and documents cross-region interpretation guidance (\S8).

\end{abstract}

\section{Introduction}\label{introduction}

The electricity demand of AI inference has moved from a rounding error to a planning constraint for cloud operators \cite{ref7,ref12}, a trajectory flagged early in the ML community\textquotesingle s accounting of deep-learning energy cost \cite{ref17}. Unlike training, inference is \emph{continuous, latency-sensitive, and geographically flexible}: the same request can often be served from any of several regions with spare capacity. Because the marginal carbon intensity of electricity --- the emissions of the generator that responds to one more MWh of demand \cite{ref1} --- differs sharply across grid regions and hours, \emph{where} and \emph{when} an inference request runs materially changes its carbon impact.

Carbon-aware computing has an established literature for batch and deferrable workloads: temporal shifting of flexible jobs \cite{ref5}, elastic scaling to track clean hours \cite{ref9}, datacenter-level carbon planning \cite{ref4,ref6}, and virtualized energy interfaces for applications \cite{ref8}. A growing body of work targets ML specifically \cite{ref13,ref14,ref15}, and geographical load balancing has long been studied for cost and, more recently, for carbon and equity \cite{ref10,ref16}. Our contribution is \emph{not} a new scheduling algorithm. It is a live systems validation of the simplest possible lever --- a per-request routing bias --- under production-like conditions, and a practical observation about \emph{which carbon signal} that lever should consume.

This distinction matters because the carbon signal a system routes on is a design choice with non-obvious consequences. WattTime exposes both an absolute MOER (lbs CO\textsubscript{2}/MWh) and a normalized signal-index (0--100 percentile of that region\textquotesingle s MOER against that region\textquotesingle s own distribution; the exact window is API-version-dependent, see \S2.2). The signal-index is an excellent instrument for the \emph{temporal} question --- "where does this region's current MOER fall within its forecast distribution over the next 24 hours?" --- which is exactly what a single-site, load-shifting controller needs. For short inference requests the current interval is the relevant one; a job spanning more than a single five-minute interval should consult the forecast rather than the instantaneous value. For the \emph{cross-region} question --- "which of these regions is cleaner right now?" --- the absolute MOER is the more natural basis, since the percentile is defined relative to each region\textquotesingle s own temporal variation.

\textbf{Contributions:}

\begin{enumerate}
\def\labelenumi{\arabic{enumi}.}
\tightlist
\item
  \emph{A live demonstration of feasibility} - that MOER signals can drive real workload migration across a live multi-region GPU testbed, as a reversible overlay on the existing production pressure router and with no observed dispatch failures (\S3, \S5.1).
\end{enumerate}

\begin{enumerate}
\def\labelenumi{\arabic{enumi}.}
\tightlist
\item
  A telemetry-calibrated energy and settled-carbon methodology: per-request energy attributed from GPU-telemetry (NVIDIA DCGM) concurrency curves, and settlement against \emph{historical} MOER, with the production pressure router as the \emph{blind} control arm (\S3--4).
\item
  Live A/B results across two real multi-region testbeds that confirm the mechanism moves load, paired with an honest read that the carbon \emph{magnitude} is a grid- and season-dependent question we quantify with a year-long historical-MOER replay, rather than longer live runs (\S5--6).
\item
  A practical observation for cross-region routing --- compare regions by absolute MOER, not the within-region percentile signal-index --- and the roadmap for the remaining joint work (\S8).
\end{enumerate}

\section{Background and related work}\label{background-and-related-work}

\subsection{Marginal vs. average emissions}\label{marginal-vs.-average-emissions}

The carbon impact of a \emph{change} in electricity consumption is governed by the marginal emissions rate --- the carbon intensity of the marginal generator --- not the grid-average rate \cite{ref1,ref19}. WattTime\textquotesingle s MOER operationalizes this as a real-time and forecast signal per grid region (typically the balancing authority); complementary flow-tracing methods underpin real-time carbon-intensity accounting for other grids and public datasets \cite{ref11} (for regions with interchange, MOER itself incorporates the marginal fuel mix of imports alongside the domestic marginal mix). For a decision that \emph{moves} load between regions or hours, marginal factors are the correct basis; average factors systematically misprice the effect of the decision \cite{ref1,ref19}. According to the Greenhouse Gas Protocol, the impact from interventions that cause a change in electricity use should be accounted for using consequential methods and marginal emissions rates \cite{ref28,ref29}. All carbon accounting in this work uses MOER.

\subsection{The signal-index is a temporal, relative instrument}\label{the-signal-index-is-a-temporal-relative-instrument}

WattTime additionally publishes a signal-index: a 0--100 value representing the \emph{percentile} of a region\textquotesingle s current MOER within that region\textquotesingle s own distribution (0 = cleanest for that region, 100 = dirtiest for that region). The exact normalization window is API-version-dependent --- legacy V2 (no longer available) compared against roughly the trailing month, whereas V3 compares against the region\textquotesingle s upcoming 24-hour forecast. These experiments use the WattTime V3 API \cite{ref26} (\texttt{co2\_moer} from \texttt{/v3/forecast} for live dispatch and \texttt{/v3/historical} for settlement), North-America MOER model version 2026-03-01 \cite{ref27} (which updated curtailment handling; \S6). This normalization is deliberately \emph{relative to each region}, which makes it ideal for a controller asking "should I run my flexible job now, or wait?" --- the question at the heart of temporal-shifting systems \cite{ref5,ref9}. The relativity that makes it excellent for \emph{within-region temporal} decisions is precisely why the absolute MOER is the better basis for \emph{cross-region spatial} decisions, as we discuss in \S5.3.

\subsection{Carbon-aware and energy-aware ML serving}\label{carbon-aware-and-energy-aware-ml-serving}

Prior ML-serving work has exploited model-variant selection and spatiotemporal scheduling to cut inference carbon \cite{ref13,ref14} and has characterized inference-cluster energy and its optimization \cite{ref15,ref16}. Measurement studies have quantified the carbon intensity of AI in cloud instances and its regional/temporal variability \cite{ref18}, established methodologies for ML carbon accounting that account for location and time-of-day effects \cite{ref20}, and characterized the broader cost of AI deployment \cite{ref12}. Closest to this work, a recent line designs carbon-aware inference serving with operational and embodied accounting from production traces \cite{ref23}, evaluates geographic LLM serving and scaling on millions of production requests \cite{ref24}, and performs per-request carbon routing under p95-latency and accuracy constraints \cite{ref25}. Our contribution is distinct along four axes those studies do not jointly cover: a live physical multi-region deployment, marginal-emissions routing, a production pressure-router control arm, and forecast-to-historical settlement. We otherwise hold the model fixed (Llama-3.1-8B-Instruct on vLLM) and isolate the effect of a \emph{routing bias alone}, measured end-to-end on live grids, so that the reported numbers are attributable to placement rather than to model or systems co-optimizations.

\subsection{What is new here}\label{what-is-new-here}

Relative to prior work, this study contributes (i) a \emph{live} rather than replay/simulated evaluation, (ii) a \emph{realistic control arm} (the production pressure router, not uniform placement), (iii) \emph{telemetry-calibrated} energy attribution and \emph{historical-MOER settlement}, and (iv) a practical percentile-vs-absolute cross-region pitfall --- a direct consequence of region-local normalization that we have nonetheless not seen documented empirically for a live production carbon signal.

\section{System design}\label{system-design}

\subsection{Routing term}\label{routing-term}

The production routing fabric (Solyx AI Grid \cite{ref22}) routes each request by a pressure weight per cell (a function of live queue depth, load, and capacity). Carbon-awareness is added \emph{multiplicatively and default-off}: each cell\textquotesingle s weight is scaled by a cleanliness multiplier

\begin{verbatim}
multiplier(cell) = 1 - w * intensity(region(cell))
\end{verbatim}

where \texttt{w} $\in [0,1]$ is the carbon weight (0 = off, i.e. the unmodified production router) and \texttt{intensity} $\in [0,1]$ is the region\textquotesingle s normalized carbon intensity. With \texttt{w\ =\ 0} the multiplier is identically 1, so the carbon term is a strict, reversible overlay on production behavior. Region identity is resolved from a verified datacenter$\rightarrow$grid region registry, not from the cloud provider\textquotesingle s (frequently incorrect) region labels.

The current implementation also sets the multiplier to 1 when a region\textquotesingle s intensity is unknown or stale --- a neutral default that, we note, can bias traffic toward a missing-data region over one with a known non-zero penalty. The safer behavior, which we recommend and are adopting, is to disable the overlay for that routing decision and revert all candidates to the pressure-only baseline rather than treat unknown as clean.

The \textbf{critical design variable} is how \texttt{intensity} is derived from the WattTime signal (\S5.3): from the \emph{absolute} MOER or from the \emph{percentile} signal-index.

\begin{figure}[htbp]
\centering
\includegraphics[width=0.98\linewidth]{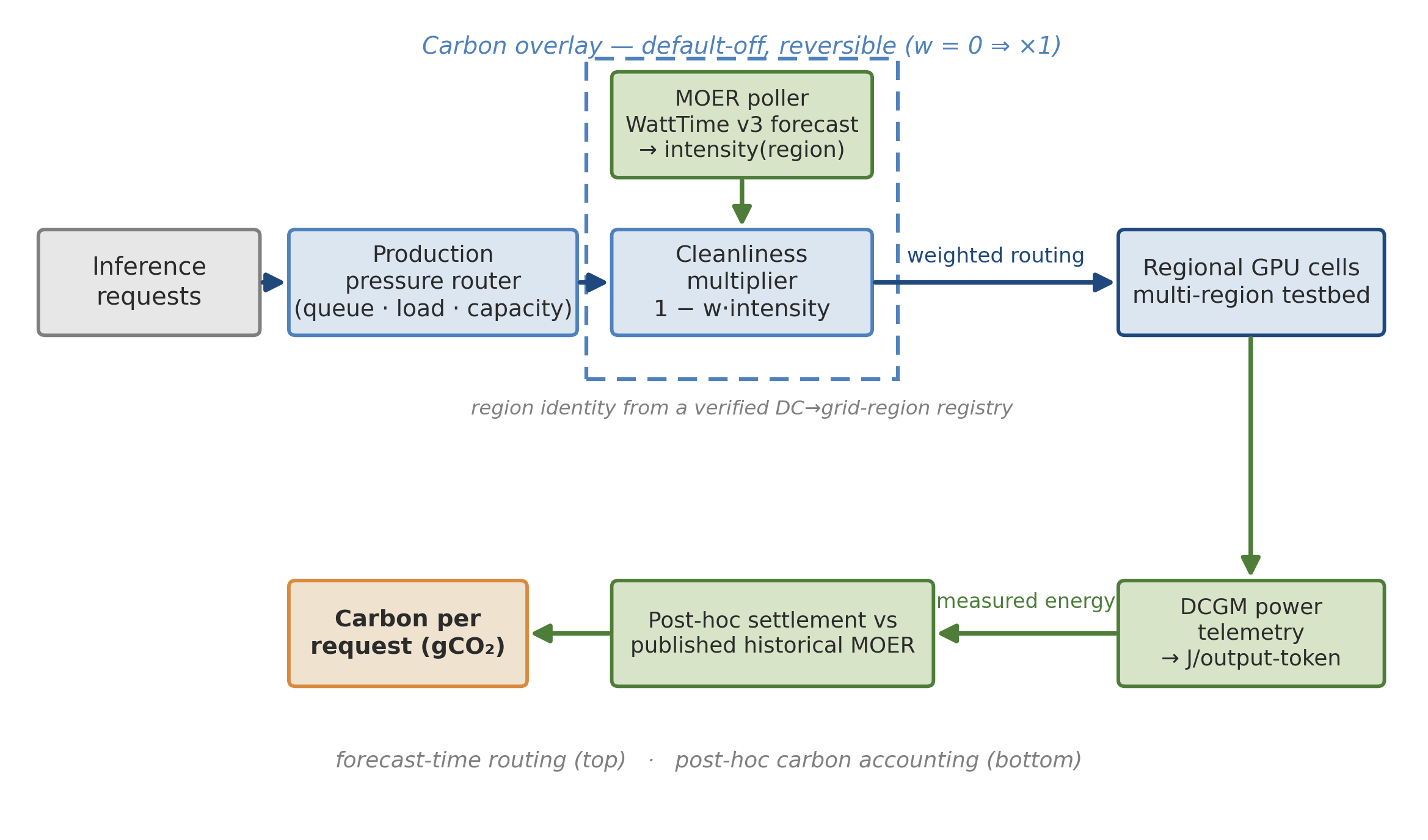}
\caption{Carbon-aware routing as a default-off, reversible overlay. Forecast MOER scales each cell\textquotesingle s pressure weight (top); energy and carbon are settled post-hoc against published historical MOER (bottom). At w = 0 the multiplier is 1 and the production router is unchanged.}
\label{fig:arch}
\end{figure}

\subsection{Energy from telemetry, not nameplate}\label{energy-from-telemetry-not-nameplate}

Per-request energy is attributed from a Phase-0 energy characterization: an offline concurrency sweep on each GPU SKU that measures GPU power via NVIDIA DCGM and fits a joules-per-output-token curve across load levels. The measured curves are steep --- energy per output token falls $\sim$30$\times$ from low to high concurrency as fixed overheads amortize:

\begin{table}[htbp]\centering\small
\caption{Per-output-token energy from the Phase-0 DCGM concurrency sweep.}\label{tab:energy}
\begin{tabular}{@{}lccc@{}}\toprule
GPU SKU & J/token @ low conc. & J/token @ sat. & Reduction\\\midrule
H100 SXM5 80\,GB & 3.37 & 0.104 & ${\sim}32\times$\\
A100 SXM4 40\,GB & 3.51 & 0.123 & ${\sim}28.5\times$\\\bottomrule
\end{tabular}\end{table}

This makes energy --- and therefore carbon --- a quantity grounded in measured device telemetry across concurrency rather than a nameplate TDP estimate. Per-request values are \emph{attributed} from these fitted curves, not integrated around each isolated request; \S7 notes the request-level factors this omits.

\begin{figure}[htbp]
\centering
\includegraphics[width=0.72\linewidth]{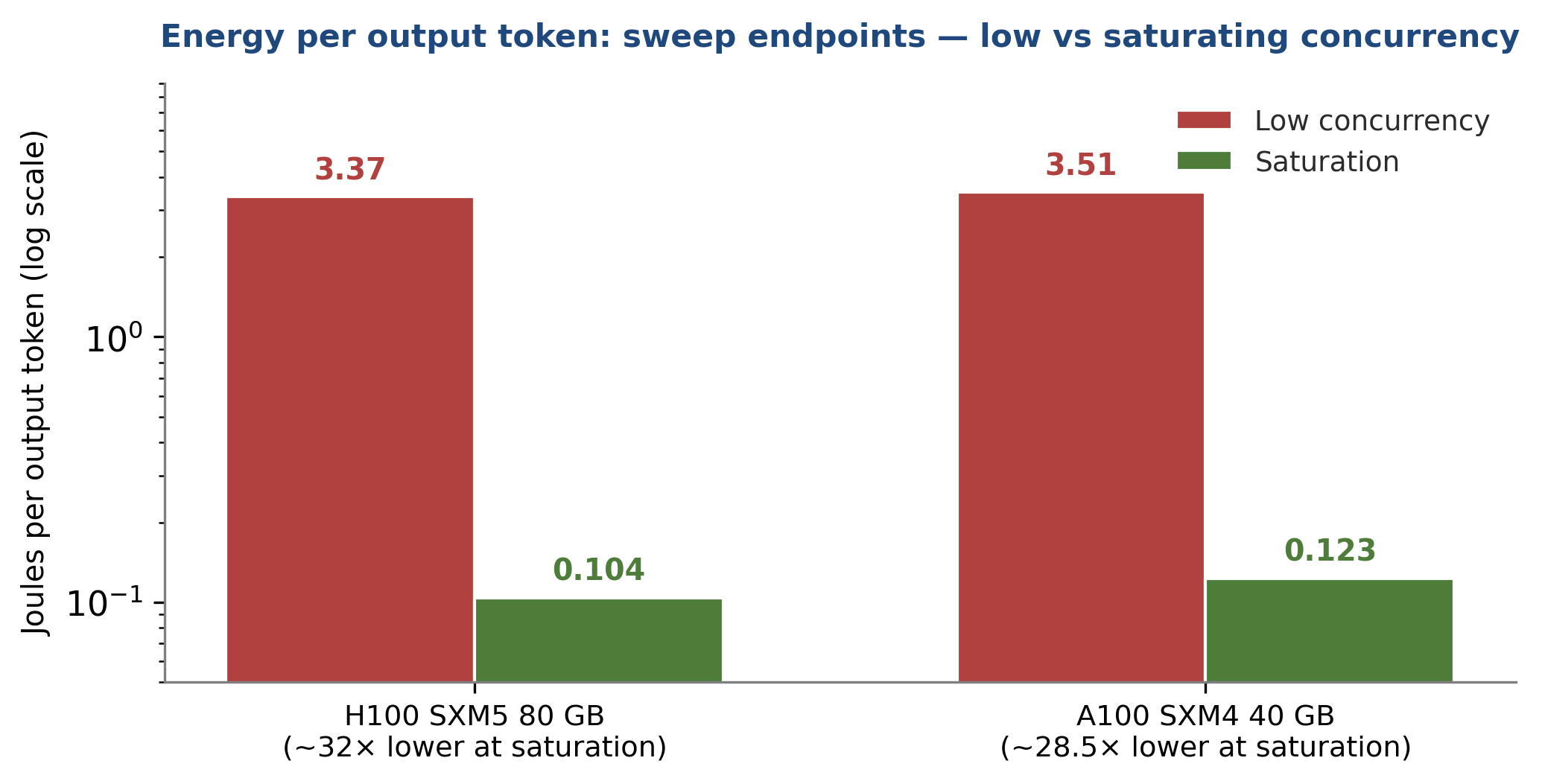}
\caption{Measured energy per output token at the two ends of the Phase-0 concurrency sweep --- low concurrency vs saturation (NVIDIA DCGM). Continuous batching amortizes fixed overhead, cutting per-token energy $\sim$28--32$\times$, so utilization --- not grid intensity alone --- governs per-request carbon.}
\label{fig:energy}
\end{figure}

\subsection{Dispatch vs. settled carbon}\label{dispatch-vs.-settled-carbon}

Routing decisions are necessarily made against the forecast MOER available at decision time. But the honest carbon attribution must use the historical MOER for the interval in which the request actually ran. The fabric records the dispatch (forecast) MOER on every request and, after the fact, settles each request against WattTime\textquotesingle s published historical MOER (5-minute resolution) \cite{ref2} before any carbon figure is quoted. Historical MOER is itself a modeled estimate that WattTime may later revise --- not a directly metered quantity --- so we treat it as the best available post-hoc reference rather than ground truth. All headline numbers in \S5 are \emph{settled}.

\section{Experimental methodology}\label{experimental-methodology}

\textbf{Workload:} A synthetic but sustained open-loop stream of Llama-3.1-8B-Instruct completions via vLLM, identical across arms, driven at a 5 arrivals/s session-initiation rate with bounded max-inflight (256). Single-turn classes emit one request per arrival while multi-turn sessions expand to several turns each, so the 48 h run produces $\approx$1.96M individual request-turns ($\approx$864k arrivals $\times$ $\approx$2.3 turns per arrival), partitioned across the five arms. Arrival-level class fractions are the workload-generator defaults --- 40\% single-turn completions, 50\% multi-turn sessions, 10\% long-context RAG --- which, at $\approx$3.5 turns per multi-turn session, correspond to $\approx$18\% / $\approx$78\% / $\approx$4\% of request-turns. Single-turn prompts are lognormal (median $\approx$245 tokens) with 64--512-token decodes; multi-turn prompts start at $\approx$181 tokens and grow $\approx$200 tokens per turn (2--16 turns, mean $\approx$3.5) with 64--320-token decodes; and the long-context class carries 2,000--4,000-token prefills with 128--512-token decodes --- the prefill-heavy tail that dominates TTFT under saturation.

\textbf{Arms:} Five routing arms run concurrently over the same workload and the same live MOER stream: \texttt{round-robin}, \texttt{off} (the production pressure router --- the \emph{blind} control), and \texttt{eco-low}/\texttt{eco-med}/\texttt{eco-high} (carbon weights 0.25/0.50/1.00). The comparison we report is eco-high vs off.

\textbf{Arm assignment and isolation:} Each request is assigned to a single arm by a seeded randomization stratified by traffic class --- requests are \emph{not} cloned across arms --- so the $\sim$1.96M request-turn count is the total across arms ($\approx$0.39M/arm), and the class mix (and thus the prompt/decode-token distribution) is balanced across arms in expectation rather than matched per request. The arms are not isolated replicas: they share the same physical GPUs, queues, and live pressure state. Consolidation by a carbon arm therefore changes the queue depth and contention the other arms experience --- a treatment-interference (SUTVA) effect we disclose rather than assume away --- so the live latency result is \emph{observational}, not a controlled per-arm measurement. Emissions are compared as per-request attributed gCO\textsubscript{2} (attributed energy $\times$ settled MOER) aggregated per arm, with shared/idle GPU power split by work share. Because the live carbon delta is small, we treat the traffic-placement result as the robust live finding and the settled-carbon delta as directional only. The \texttt{off} and \texttt{eco-high} arms use the same underlying pressure router and differ in their assigned carbon multiplier, so their contrast reflects enabling the carbon-aware policy within this shared-resource environment. Because the arms influence one another\textquotesingle s queue and contention state, the contrast is not an isolated per-arm causal effect.

\textbf{Reading the live runs:} These two-site runs are a \emph{feasibility and systems} check, not a powered measurement: at only a couple of diurnal cycles per testbed they cannot support confidence intervals. We therefore report what was observed --- whether the signal moved load, the observed latency and dispatch-failure behavior, and whether attributed emissions settled in the intended direction --- and reserve statistical inference (confidence intervals, seasonal resolution) for the historical replay (\S6). Every request\textquotesingle s carbon is settled against historical MOER before any figure is quoted.

\textbf{Testbeds:} Each is a two-region, two-GPU \emph{testbed cell} / live production-cluster slice, not a fleet; we reserve "fleet" for the 19-region CONUS replay (\S6).

\begin{table}[htbp]\centering\footnotesize
\caption{Live two-region GPU testbeds.}\label{tab:fleet}
\begin{tabular}{@{}l l p{3.2cm} p{3.2cm} l@{}}\toprule
GPU & Count/SKU & Clean-leaning region & Baseline region & Duration\\\midrule
A100 & 2$\times$ A100-SXM4-40 GB & AZPS --- Arizona, solar-heavy & PJM-DC (Dominion / DOM) --- Ashburn, N. Virginia & $\sim$48 h (2 cycles)\\
H100 & 2$\times$ H100-SXM5-80 GB & ERCOT North-Central --- Dallas, wind/solar, wide-swing & SOCO --- Atlanta, narrow fossil baseline & Terminated early (see \S5.3, \S5.5)\\
\bottomrule\end{tabular}\end{table}

Region selection was constrained by real cloud H100/A100 availability; notably, no H100-class capacity was available in any deep-curtailment western (CAISO) region during the study window (\S5.5). Note that the live PJM node here is PJM-DC (Dominion / DOM, Ashburn), a different PJM subregion from the PJM-Chicago (ComEd, northern Illinois) node used in the replay fleet (\S6); PJM subregions differ materially in marginal fuel mix, so the two are not interchangeable.

\section{Results}\label{results}

\subsection{Systems viability}\label{systems-viability}

The fabric ran unattended on the live testbeds for multi-day spans, sustaining approximately 1.96 million request-turns with zero dispatch errors, live 5-minute MOER polling, atomic checkpointing, and post-hoc settlement. The carbon overlay is a strict multiplicative layer on the production router and is fully reversible (\texttt{w=0}). This establishes the primary systems claim: carbon-aware routing is deployable on a production inference fabric with no observed dispatch failures over the test window, and with a clean off-switch. Reliability dimensions beyond dispatch success --- request-completion/timeout rates, SLO attainment, failover --- were not separately measured and are not claimed here.

\subsection{The mechanism moves real load (A100 $\cdot$ AZPS + PJM-DC)}\label{the-mechanism-moves-real-load-a100-azps-pjm-dc}

Over $\sim$48 hours ($\sim$1.96M logged request-turns total across arms --- $\approx$864k session/single-turn arrivals at 5/s, expanded by multi-turn sessions --- and 0 dispatch errors), the eco-high arm settled 1.45\% lower attributed GPU operational emissions than the off (blind pressure-router) arm. On latency: under the deliberately saturating workload, the observed p95 was 18.7 s for the off arm and 20.9 s for eco-high, an 11.7\% increase (the high absolute p95 reflects the deliberately saturating open-loop load and the 2,000--4,000-token prefills in the long-context class; see \S4). Because the routing arms shared physical GPUs, queues, and pressure state, this is an observational within-run comparison rather than an isolated causal estimate; the experiment was not designed or powered to establish a generalizable latency effect or formal SLO non-inferiority. We report the emissions figure as an observed, non-generalizable descriptive result, not an estimate of expected savings: a single $\sim$48-hour window, one grid pair, roughly two diurnal cycles, and highly autocorrelated MOER conditions carry little statistical power for a magnitude claim --- even though the request count is large. Millions of requests inside one weather-and-grid period are not millions of independent carbon observations.

The load-bearing result of this run is directional: a WattTime signal drove real inference load from one live region to another, and settled emissions moved in the intended direction, end-to-end, on a live production router. Shifting load away from an already-optimized production baseline --- not a round-robin straw man --- under a deliberately saturating workload is the non-trivial part; the small emissions delta is expected for a region pair whose MOER distributions only moderately diverge in a low-curtailment week, and is not the lever\textquotesingle s ceiling. Across the full five-arm slider the response is monotonic (round-robin $\rightarrow$ off $\rightarrow$ eco-low $\rightarrow$ eco-med $\rightarrow$ eco-high): the clean-region (AZPS) traffic share rises 50 $\rightarrow$ 56 $\rightarrow$ 57 $\rightarrow$ 58 $\rightarrow$ 63\%, settled gCO\textsubscript{2}/req falls, and p95 latency rises --- we report eco-high vs off as the extreme contrast, but the intermediate tiers move as expected, not selectively. Characterizing the actual magnitude across grids and seasons is the job of the historical replay (\S6), which supplies the confidence intervals and seasonal structure a single live run cannot.

\subsection{An interpretation note: rank regions by absolute MOER (H100 $\cdot$ ERCOT-NC + SOCO)}\label{an-interpretation-note-rank-regions-by-absolute-moer-h100-ercot-nc-soco}

We initially derived \texttt{intensity} from WattTime\textquotesingle s signal-index (percentile). On the H100 testbed the carbon arm moved load but tracked each region\textquotesingle s \emph{percentile} rather than its \emph{absolute} emissions --- so it did not reliably prefer the cleaner region. The reason is a straightforward property of the signal:

\begin{quote}
The signal-index is a percentile relative to each region\textquotesingle s own distribution. When two regions have very different MOER distributions, equal percentiles do not imply equal absolute emissions --- and the percentile ranking can invert the absolute ranking.
\end{quote}

Concretely, Dallas (ERCOT-NC) is a wide-swing grid (MOER historically ranging $\sim$0--1389 lbs/MWh); Atlanta (SOCO) is narrow ($\sim$1135--1390). A Dallas reading of $\sim$1253 lbs/MWh sits at a \emph{high} percentile for Dallas, while an Atlanta reading of $\sim$1330 lbs/MWh sits at a comparatively \emph{lower} percentile for Atlanta --- even though Dallas is absolutely cleaner (1253 \textless{} 1330). A percentile-driven router therefore sends load to the absolutely-\emph{dirtier} region. The signal-index is doing exactly what it was designed to do (flag each region against its own distribution); it is simply the wrong metric for a spatial decision.

\begin{figure}[htbp]
\centering
\includegraphics[width=0.95\linewidth]{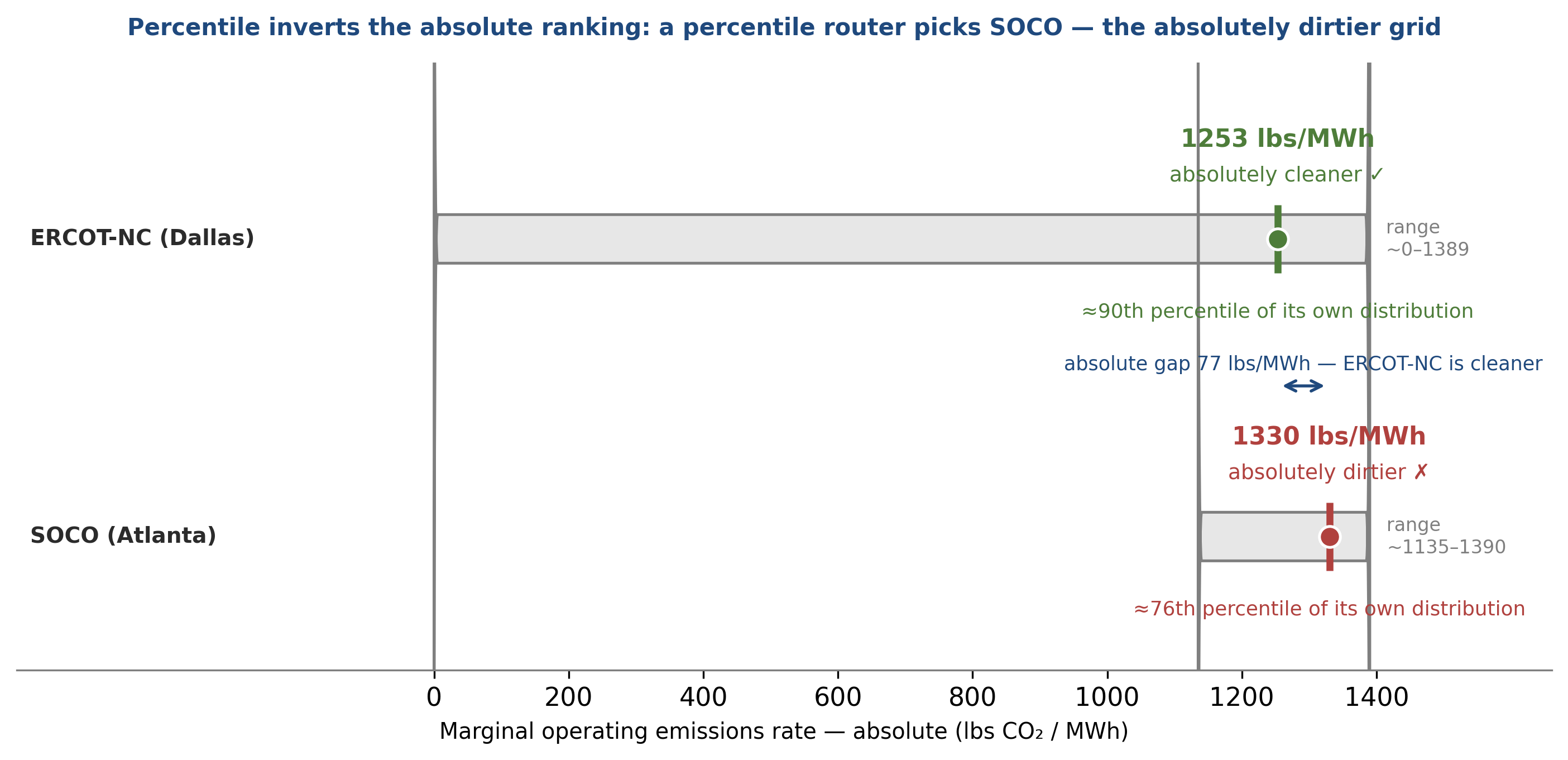}
\caption{Why cross-region comparisons should use absolute MOER, not the percentile signal-index. On a shared absolute scale ERCOT-NC (1253 lbs/MWh) is cleaner than SOCO (1330), yet sits at a higher percentile within its own wide-swing distribution- so a percentile-driven router inverts the ranking and picks the dirtier grid.}
\label{fig:inversion}
\end{figure}

\textbf{Fix:} Deriving \texttt{intensity} from absolute MOER, normalized by a fixed reference scale, aligns routing with the absolute ranking: the carbon arm prefers the absolutely-cleaner region --- modestly when regions are close, strongly when one region curtails. We verified the corrected direction in live telemetry (the carbon arm\textquotesingle s share of the cleaner region rose above the blind arm\textquotesingle s, tracking the sign of the absolute MOER gap). It is a one-line change in \emph{how the signal is interpreted} --- absolute MOER for spatial comparisons, the percentile for within-region timing. The absolute MOER (historical and forecast) also serves the within-region timing question, and is what licensed users typically route on; the signal-index is freely available as a shifting signal but does not expose the absolute values needed to quantify a carbon reduction.

\textbf{A note on the signal-index:} The signal-index is built for the temporal question and excels at it; the observation here is simply that cross-region \emph{comparisons} are better served by the absolute MOER. As multi-region inference grows, that is a useful and easily-documented clarification for signal-index consumers (\S8).

\subsection{Energy characterization}\label{energy-characterization}

The Phase-0 curves (\S3.2) show energy-per-output-token falling $\sim$30$\times$ from low to saturating concurrency on both H100 and A100. Two implications: (i) carbon savings from \emph{routing} compound with, and can be dominated by, \emph{utilization} --- an underloaded clean region can emit more per token than a saturated dirtier one; and (ii) any credible inference-carbon claim must measure energy at the operating concurrency, not from nameplate TDP, consistent with calls for systematic energy-and-carbon reporting in ML \cite{ref21}.

\subsection{Magnitude is grid- and season-dependent}\label{magnitude-is-grid--and-season-dependent}

Live forecasts during the H100 window explained the muted signal: in mid-July, Dallas (ERCOT-NC) deep curtailment did not appear in the forecast; its forecast MOER stayed pinned in a narrow $\sim$1187--1279 lbs/MWh band rather than its full $\sim$0--1389 range --- under high air-conditioning demand and weak wind. The usable cross-region spread (peaking $\sim$+180 lbs/MWh) came largely from \emph{SOCO rising} into its afternoon peak, not from Dallas dipping, and even inverted overnight. The single most curtailment-rich scenario (CAISO) was unavailable on H100-class hardware for the entire window. Takeaway: the \emph{mechanism} (absolute-MOER routing) is validated and directionally correct; the \emph{magnitude} is a strong function of grid pair, curtailment depth, and season --- a breadth question best answered by historical-MOER replay across grids and seasons (\S6), not by any single two-site live run.

\section{Historical replay: the magnitude question}\label{historical-replay-the-magnitude-question}

The live runs establish feasibility but cannot size the effect (\S5.5). To answer how much carbon-aware routing actually saves, we replay a structured synthetic inference workload --- with diurnal, weekday, seasonal and multi-turn-session behavior specified below --- against a full year of historical MOER across a grid-diverse CONUS fleet, the instrument a two-site live run cannot be. Because the replay dispatches against the historical MOER series itself rather than against a forecast, it is a perfect-foresight scenario: the reductions it reports are an upper bound --- the full available opportunity --- rather than what forecast-driven dispatch would realize in operation. Forecast error would move any deployed system below this ceiling. ("Realistic" is reserved for a future run against an actual production trace.)

\textbf{Setup:} The dataset contains 8,761 hourly timestamps bounding 8,760 consecutive one-hour intervals from 2025-07-06 16:00 UTC through 2026-07-06 16:00 UTC (a full 365-day year). Each interval uses the MOER observation at its starting boundary (timestamp \texttt{t} covers \texttt{{[}t,\ t+1\ h)}); the 8,761st timestamp is the year\textquotesingle s closing boundary. MOER is provided for 19 CONUS grid regions, with zero-MOER intervals retained as real zeros rather than dropped (\S curtailment treatment below). Requests are placed by the same carbon-aware overlay evaluated live (\S3.1), with energy represented using a homogeneous H100-equivalent near-saturation energy scale of $\sim$0.104 J/output-token (the replay does not dynamically apply the full concurrency curve --- \S3.2, \S5.4) and carbon settled against the historical MOER series. Confidence intervals come from a 5-day block bootstrap (10,000 resamples) \cite{ref3}, which preserves the temporal autocorrelation the live runs could not. The evaluated configuration includes a modeled multi-turn session layer --- a fraction of load pinned to where each conversation began --- which \emph{lags} the carbon-optimal placement and so makes the headline conservative: it lowers the idealized number, never inflates it.

Headline\textbf{:} Magnitude is a policy-and-treatment frontier, not a single number. The replay indicates reductions from $\sim$23\% (22.7\%, 95\% block-bootstrap CI 21.9--23.5\%) under a gentle soft-multiplier policy matching the live production overlay\textquotesingle s form (the \S3.1 weighting at carbon weight \texttt{w=1}, under the same per-region capacity cap) up to 34--51\% across the aggressive cleanest-first dispatch and alternative treatments of the zero-MOER curtailment intervals (the $\alpha$-sensitivity below). In the primary configuration - capacity-capped cleanest-first dispatch at a 50\%-per-site cap, with the multi-turn session layer - carbon-aware placement reduces modeled GPU-attributable operational emissions by a central 50.9\% versus round-robin and 51.0\% versus the modeled carbon-blind load balancer (which is marginally dirtier than round-robin here; see baseline robustness below) (95\% block-bootstrap CI 48.5--53.3\%). That CI is \emph{conditional on the modeled fleet, routing policy, workload and zero-MOER-credit assumptions}: it captures temporal sampling error in the chosen historical series only, not uncertainty from fleet composition, physical headroom, energy attribution, or the WattTime model. The 50\%-per-site cap is itself a permissive concentration assumption: with 19 regions under uniform load, baseline share is $\sim$5.3\% per region, so the cap lets a single region absorb roughly 9.5$\times$ its baseline traffic --- more concentration than most operational fleets could accept without breaching queue, memory or latency limits. The gentle soft-multiplier arm ($\sim$23\%), which mirrors the live overlay's form, is therefore the more operationally conservative reference point, and the 50.9\% should be read as the capacity-capped upper bound of the frontier. The 50.9\% already absorbs the session penalty --- pinning $\sim$40\% of load to its origin lowers an idealized freely-placeable $\sim$54\% to 50.9\%, so on that axis it is conservative. Even the frontier\textquotesingle s low end is an order of magnitude above the 1.45\% observed live, with statistical backing the single live run lacks --- but it is the output of one specified configuration (below), not a universal estimate.

\textbf{Baseline robustness:} The figure is quoted versus round-robin, which differs from the production pressure router used as the live control (\S5.2). To probe whether that matters, the replay also carries a \emph{carbon-blind load-balancing} arm (serve-local, spill on saturation) --- a replay analog of the pressure router. The modeled load balancer produced approximately 0.3\% higher attributed emissions than round-robin, so the carbon-aware policy reduced emissions by 50.9\% relative to round-robin and 51.0\% relative to the modeled load balancer. That the two baselines land so close suggests - but does not establish - that the headline is relatively insensitive to the particular carbon-blind baseline. Two caveats keep this from being conclusive: the arm balances by modeled regional demand rather than the pressure router\textquotesingle s queue/TTFT signals, and the modeled demand across these 19 regions is close to uniform (assumptions below), so a carbon-blind balancer has little skew to exploit and sits near round-robin almost by construction. A like-for-like replay against the production router\textquotesingle s actual signals is future work; for now we claim only that the result does not obviously depend on choosing round-robin over a carbon-blind balancer.

\textbf{Interval-width sensitivity:} The CI widens modestly with the bootstrap block length --- {[}49.4, 52.4{]} at 1-day, {[}48.5, 53.3{]} at 5-day, {[}48.3, 53.6{]} at 7-day, {[}47.6, 54.4{]} at 14-day --- as longer blocks absorb weekly weather/grid autocorrelation. The point estimate is unchanged and every interval stays far from zero; we use five days as the primary specification and report the 1-, 7-, and 14-day blocks as sensitivity analyses.

\textbf{What drives it:} We decompose against a fixed counterfactual, all three sharing the equal-capacity fleet and the 50\%-per-site cap: \emph{baseline} = round-robin (even split); \emph{static} = the same capacity-capped cleanest-first fill applied to each region\textquotesingle s \textbf{annual-mean} MOER, held fixed all year (so routing shares are determined by that fill, not by hourly values); \emph{dynamic} = the identical fill re-solved each hour on the current MOER. The static annual-mean-MOER policy supplies $\sim$60\% of the total reduction and the dynamic increment (dynamic $-$ static) the remaining $\sim$40\% - in absolute terms $\approx$31.6 and $\approx$22.4 percentage points of the 54.0\% placement reduction, respectively (these are shares of the reduction, not standalone reduction percentages). Both stages consume WattTime data - static uses the annual means, dynamic the hourly series - so this isolates the value of the real-time signal specifically, not of WattTime data overall. (Note this is \emph{allocation}, not physical siting: capacity is equal across regions; nothing is preferentially built in cleaner grids.) That $\sim$40\% comes from following curtailment hour-by-hour is the result most specific to the real-time product.

\begin{figure}[htbp]
\centering
\includegraphics[width=0.92\linewidth]{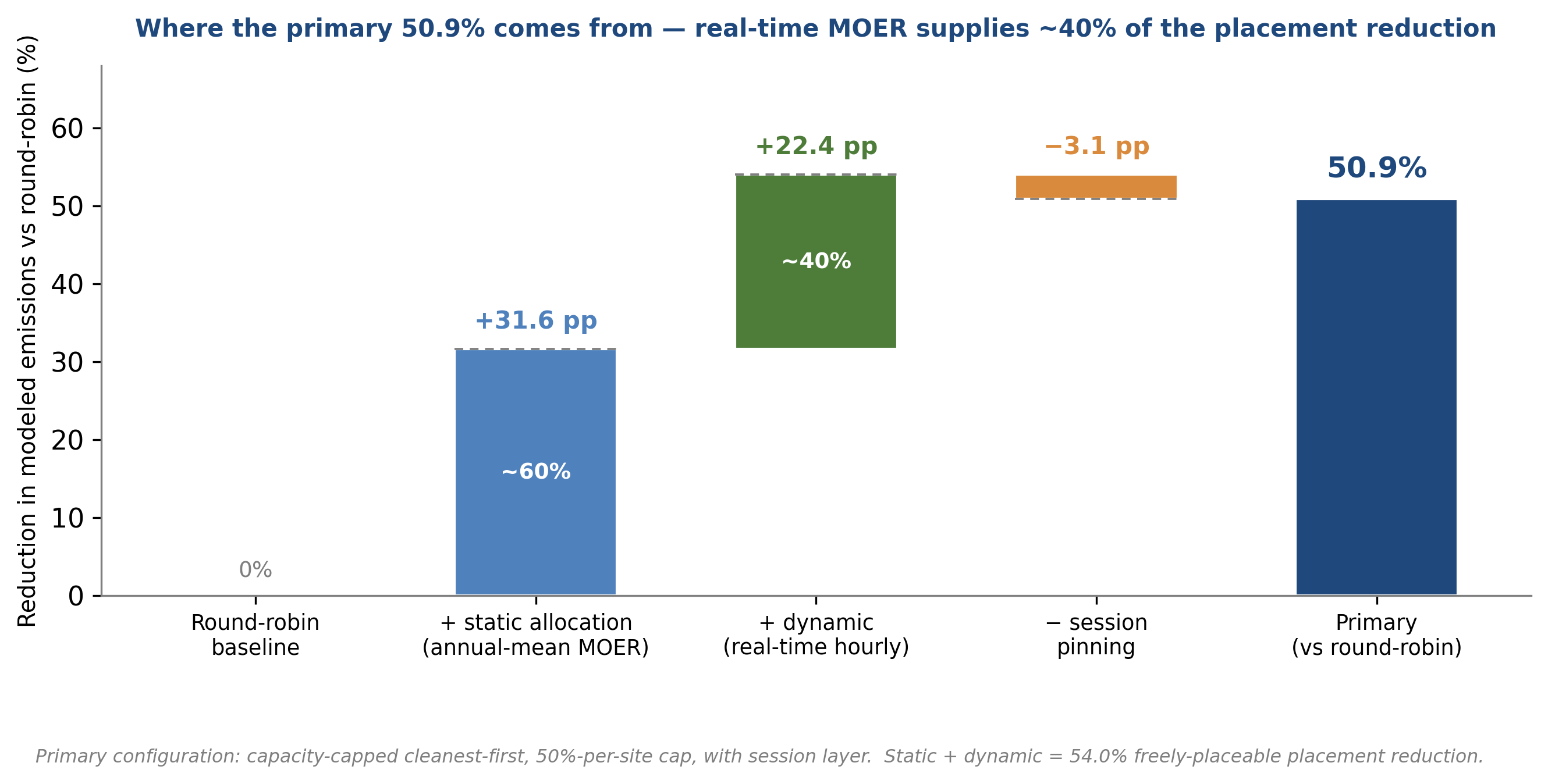}
\caption{Decomposition of the primary 50.9\% reduction (vs round-robin). Static annual-mean allocation and the dynamic real-time signal supply $\approx$31.6 and $\approx$22.4 percentage points of the 54.0\% freely-placeable placement reduction; the modeled session layer costs $\approx$3 points. Allocation across an equal-capacity fleet, not physical siting.}
\label{fig:decomp}
\end{figure}

\textbf{Seasonal and diurnal structure:} The reduction tracks curtailment: spring (MAM) reaches 63.7\%, and the diurnal profile peaks at 69.4\% at the solar-generation peak, where routing into a region spilling zero-carbon power is most valuable. This concentration is expected --- and it makes explicit that the headline is materially driven by the zero-MOER curtailment intervals; the curtailment treatment and the zero-MOER-credit sensitivity below bound exactly how much of the result depends on crediting them.

\textbf{Curtailment (zero-MOER) treatment:} A WattTime \texttt{co2\_moer} value of exactly \texttt{0.0} is retained as a real zero-carbon renewable-curtailment reading, not dropped as a sentinel. A zero MOER represents the modeled marginal operating emissions rate for incremental load while addressable renewable curtailment remains available. Stated precisely: for the model version and data export used here, WattTime confirmed that numeric \texttt{0.0} observations represent modeled renewable-curtailment intervals, while genuinely missing observations are represented as absent points --- so a real zero is technically distinguishable from a gap in the returned dataset (personal communication, WattTime, 2026-07; the North-America MOER model version 2026-03-01 \cite{ref27} upgraded this curtailment detection model which indicates more frequent zero-MOER values). Retaining the zeros lowers, e.g., CAISO-North\textquotesingle s annual-mean MOER from $\sim$942 to $\sim$732 lbs/MWh, and the zeros concentrate in the midday solar window (CAISO-North: 73--91\% of midday points), which is why the diurnal and spring carbon reduction peaks are high. As a data-cleaning sensitivity (not a policy counterfactual), simply \emph{dropping} the zero-valued timestamps instead yields $\sim$24.5\% carbon reduction rather than $\sim$51\%. Note these are different operations: dropping removes those hours from the demand-weighted evaluation and changes the temporal sample, whereas the $\alpha$-sensitivity (next) retains the same workload and hours but assigns each zero interval a conservative non-zero rate. Because the result depends on this handling, we treat zero-crediting as a pre-specified sensitivity (next); the drop-zeros figure is reported only as data-cleaning context.

\textbf{Zero-MOER-credit sensitivity (the $\alpha$-sweep):} This varies \emph{how much emissions credit the zero-MOER curtailment intervals receive}, not physical capacity - it is a curtailment-\emph{value} sensitivity, not a headroom sweep. Each \texttt{0.0} is repriced at \texttt{$\alpha$\ $\times$\ (region\ median\ non-zero\ MOER)} and the replay re-run end-to-end: \texttt{$\alpha$=0} (full zero-carbon) gives $\sim$51\%, \texttt{$\alpha$=1} (curtailment credited as worth nothing - each zero repriced at that region's median non-zero MOER) still gives $\sim$34\%, which is why we report the reduction as a 34--51\% range under alternative treatment of the zero-MOER intervals rather than a point estimate. A separate, genuinely physical headroom constraint - how many curtailed MW a region can actually absorb, or a per-region capacity cap in MW - is \emph{not} modeled here; per-site curtailment-MW data would let us add it as a capacity constraint alongside $\alpha$ (a two-dimensional savings = f(headroom, $\alpha$) analysis), and is future work (\S8). WattTime's curtailment treatment assumes that the volume of addressable curtailed renewable energy exceeds the volume of load being shifted. Where that holds, the applicable modeled MOER is zero; once curtailment is exhausted, the applicable MOER becomes positive. The $\alpha$-sweep therefore explores a pragmatic sensitivity range from zero to the region's median non-zero MOER; it should not be interpreted as a physical bound on the post-curtailment marginal rate.

\textbf{Energy attribution vs. \S5.4:} The replay uses a fixed per-token energy (regions modeled near saturation, where the Phase-0 curve flattens toward $\sim$0.104 J/token) and therefore isolates the effect of MOER-aware placement from utilization-dependent energy effects. In an operational deployment, routing changes regional concurrency, batching efficiency and per-token energy; because those shifts are \emph{induced by} the MOER-aware policy - it concentrates traffic in low-MOER regions, which may become more (or, under congestion/poor batching, less) efficient while regions losing traffic slide down the curve - they could increase or decrease the reported percentage reduction, not merely the absolute grams. Given the $\sim$30$\times$ spread in per-token energy across the concurrency curve, this is potentially material. Coupling the Phase-0 concurrency curves directly to the replay is therefore necessary to determine the net effect, and is future work (\S8).

\textbf{Configuration and assumptions:} The 50.9\% is the output of one specified configuration; the primary assumptions, and what is \emph{not} modeled, are:

\begin{table}[htbp]\centering\footnotesize
\caption{Historical-replay configuration.}\label{tab:replay}
\begin{tabular}{@{}>{\raggedright\arraybackslash}p{3.4cm} >{\raggedright\arraybackslash}p{11.2cm}@{}}\toprule
Regions & 19 CONUS grid regions / subregions (CAISO-North/SanDiego/SanBernardino, LDWP, AZPS, SRP, PNM, BPA, ERCOT-Panhandle/Coast, SPP-Sioux, MISO-N.Dakota/Indianapolis, NYISO-NYC, ISONE-NEMA, PJM-Chicago, DUK, SOCO, FPL)\\\addlinespace
Signal & WattTime V3 \texttt{co2\_moer}, model 2026-03-01 (V3 is the API/schema version, not the MOER data-model version); 8,760 hourly intervals (8,761 boundary timestamps), 2025-07-06 $\rightarrow$ 2026-07-06 UTC\\\addlinespace
Per-region capacity & Uniform: each region absorbs up to 50\% of instantaneous fleet demand (the cap acts as a dispatch-concentration limit and provisional capacity proxy --- a latency-SLO control only once cross-region latency is modeled); no per-region GPU-count heterogeneity\\\addlinespace
Energy / hardware & Fixed per-token energy, a homogeneous H100-equivalent near-saturation scale ($\sim$0.104 J/token; the A100 endpoint is $\sim$0.123); concurrency-dependent J/token shifts (\S5.4) not modeled. Under this homogeneous fixed-energy assumption the percentage reduction is independent of the common J/token scale; regional hardware heterogeneity or routing-induced utilization changes could alter the percentage if efficiency varies systematically with placement (\S5.4)\\\addlinespace
Workload & Structured synthetic: diurnal (peak 15:00 local, amp 0.4) $\times$ weekday (weekend 0.75) $\times$ seasonal (peak mid-July, 1.15); origin-demand skew is illustrative and $\approx$uniform across these 19 regions\\\addlinespace
Routing (carbon-aware) & Capacity-capped cleanest-first: fill lowest-MOER region to the 50\% cap, then the next (\texttt{greedyCapDist})\\\addlinespace
Routing (soft policy) & Even share $\times$ \texttt{(1\ $-$\ w$\cdot$intensity)}, normalized, then clipped to the same 50\% per-region cap and waterfilled (\texttt{softCapDist}); \texttt{w=1} gives the $\sim$23\% figure\\\addlinespace
Baselines & Round-robin (even split) and carbon-blind serve-local/spill load balancing\\\addlinespace
Sessions & 40\% of load multi-turn, pinned to origin, mean life 2 h (EMA stickiness)\\\addlinespace
Scope of "emissions" & GPU-attributable operational CO\textsubscript{2} only (energy $\times$ MOER); excludes PUE/overhead, networking, and embodied carbon\\\addlinespace
Not modeled & Cross-region latency and data-transfer/egress cost; per-region energy heterogeneity and concurrency-dependent J/token; vLLM prefix-cache (APC/RadixAttention) invalidation on cross-region dispatch; nodal (vs zonal) deliverability of curtailed power; physical curtailment-MW headroom\\
\bottomrule\end{tabular}\end{table}

The MOER series underlying this replay is licensed from WattTime and cannot be redistributed; it is obtainable directly from WattTime under their standard terms (V3 /v3/historical, North-America MOER model 2026-03-01, the region list and UTC window given above). The routing and placement logic is the subject of pending patent applications and is not released. What we specify in full is the configuration above --- fleet, regions, dispatch policy, capacity cap, session model, energy attribution and the zero-MOER-credit sweep --- which, together with a licensed MOER series, is sufficient to reconstruct this study independently. Sensitivity to session-pinning, fleet composition, and the demand model --- beyond the zero-MOER-credit $\alpha$-sensitivity already reported --- remains partly future work (\S8).

\section{Discussion and limitations}\label{discussion-and-limitations}

\begin{itemize}
\tightlist
\item
  \textbf{Feasibility, not a percentage:} The live runs are point demonstrations on specific grid pairs, models, and weeks, without the sample size for confidence intervals. They establish that the mechanism works end-to-end and moves load on the signal --- not a generalizable "carbon-aware routing saves X\%" from the live runs alone --- the historical replay (\S6) characterizes potential magnitude under the stated assumptions, with conditional confidence intervals.
\item
  \textbf{Two-region testbeds under-sample the lever:} Spatial carbon arbitrage grows with the number and diversity of regions and with capacity headroom to actually shift load. Two co-available cloud regions is a floor, not a ceiling.
\item
  \textbf{Single model, synthetic load:} We fixed Llama-3.1-8B to isolate routing; production traffic mixes model sizes and bursty arrivals. The energy curves (\S5.4) suggest utilization interacts strongly with any routing policy.
\item
  \textbf{Prefix-cache locality is not modeled:} For single-turn or unpinned requests we assume independent prompt processing. In production vLLM, Automatic Prefix Caching (APC / RadixAttention) lets requests share common prompt prefixes; cross-region dispatch of an unpinned (or unpinned multi-turn) request breaks that locality, forcing a destination-side cache miss and full prefill recomputation that raises both TTFT and prefill energy. Quantifying the interaction between cross-region carbon routing and prefix-cache invalidation remains future work.
\item
  \textbf{The H100 run was short:} The absolute-MOER direction was confirmed in live telemetry, but we did not run it long enough to quote a settled figure - and, given the mid-July no-curtailment forecast, did not try to.
\item
  \textbf{Latency tradeoff:} Eco-high showed an observed 11.7\% increase in p95 latency (18.7 s $\rightarrow$ 20.9 s) during the 48-hour saturating test. No formal SLO threshold was pre-specified, and shared-resource interference prevents interpreting this as an isolated treatment effect. A controlled replication with isolated resources or an interference-aware analysis is needed to estimate a generalizable latency--carbon tradeoff.
\item
  \textbf{Facility overhead is excluded:} All emissions here are GPU-attributable (\S6); PUE is not modeled. Because PUE varies with regional climate and cooling technology - a hot-summer desert site can run materially higher than a northern one - routing toward a lower-MOER region could offset part of the IT-level saving through higher facility overhead, or amplify it where the cleaner region is also cooler. Incorporating site-specific real-time PUE would refine total facility carbon attribution.
\end{itemize}

None of these undercut the load-bearing claim: \textbf{signal-driven workload migration is feasible on a production fabric today} - with the practical corollary that cross-region routing should rank by absolute MOER.

\section{Future work}\label{future-work}

With potential magnitude initially characterized under the replay assumptions (\S6), the remaining work is narrower and concrete.

\textbf{Add a physical headroom constraint:} The 34--51\% range is a \emph{zero-MOER-credit} sensitivity (\S6), not a capacity model. The next step is to add a genuinely physical constraint --- per-region curtailment-MW (or a regional MW/percentage cap) that bounds how much load a curtailing region can actually absorb at zero carbon --- and report a two-dimensional \texttt{savings\ =\ f(headroom,\ $\alpha$)} surface rather than the $\alpha$ range alone.

\textbf{Broader live validation:} The live testbeds covered two-region pairs in low-curtailment windows; a focused multi-season live campaign across a deep-curtailment pair (e.g. CAISO / ERCOT) would connect the feasibility and magnitude halves directly on hardware.

\textbf{Operational best practices:} Harden the overlay for production: on \textbf{stale or unknown} intensity, disable the carbon term for that routing decision and revert all candidates to the pressure-only baseline rather than treat unknown as clean (\S3.1); and couple the router\textquotesingle s concurrency-dependent per-token energy (\S5.4) and prefix-cache locality (\S7) into the placement decision so utilization and cache effects are accounted for, not assumed away.

\textbf{Absolute-first cross-region guidance (with WattTime):} Co-author concise, public guidance (ideally an SDK helper or documentation note) for signal-index consumers making \emph{spatial} decisions: use absolute MOER when \emph{comparing regions}, and reserve the signal-index for \emph{within-region temporal} shifting. We would contribute the \S5.3 case study as the motivating example --- a small, useful clarification for a broad class of WattTime consumers.

Settlement methodology is itself part of the contribution: reconciling dispatch-time forecast MOER against published historical MOER is exactly the kind of accounting this paper documents carefully, and it underpins every carbon figure reported here.

\section{Conclusion}\label{conclusion}

Signal-driven workload migration is feasible today: WattTime MOER signals can steer live inference across a production multi-region GPU testbed as a reversible overlay with no observed dispatch failures over the test window. Our live runs demonstrate that feasibility and settle emissions in the intended direction; the potential \emph{magnitude} is characterized by the historical replay a single live A/B cannot support: in the primary replay configuration, across a grid-diverse CONUS fleet, carbon-aware placement cuts modeled GPU-attributable operational emissions by a central 50.9\% versus round-robin (51.0\% versus a modeled carbon-blind load balancer, $\sim$0.3\% dirtier here) (95\% block-bootstrap CI 48.5--53.3\%, conditional on the modeled assumptions; $\sim$23--51\% across policies and the zero-MOER-credit sensitivity); before session pinning, the real-time WattTime MOER signal contributes roughly 40\% of the 54.0\% placement reduction ($\approx$22.4 points) beyond the static annual-mean-MOER policy. Along the way we note a practical point for anyone routing across regions --- compare by absolute MOER, and reserve the percentile signal-index for within-region timing. Together, the live validation and the historical replay provide complementary evidence of operational feasibility and of potential magnitude.

\section*{Acknowledgments}\label{acknowledgments}

The authors thank Lucy Matthews and Geoff Hancock of WattTime for technical review of this manuscript, and in particular for guidance on marginal-emissions terminology, the definition and correct interpretation of the signal-index, and the treatment of renewable curtailment in the MOER series. The marginal-emissions data used in this study were licensed from WattTime. Responsibility for the analysis, results, and any remaining errors rests with the authors.

\end{document}